\documentclass[aps,twocolumn,showpacs,superscriptaddress]{revtex4}

\usepackage{graphics}
\usepackage{epsfig}
\usepackage{graphicx}
\usepackage{amsmath,amssymb,amsthm,bbm,latexsym}    
\usepackage{amsthm} 
\usepackage{amsbsy} 

\newcommand{\ndash}{\textendash}
\newcommand{\mdash}{\textemdash}
\newcommand{\ie}{{\it i.e. }}

\newcommand{\etal}{{\it et al. }}
\newcommand{\eg}{{\it e.g. }}

\newcommand{\mi}{\mathrm{i}}

\newcommand{\ket}[1]{\left|#1\right\rangle} 

\newcommand{\C}{\mathbbm C} 

\newcommand{\ucla}{\affiliation{Electrical Engineering Department, University of California at Los Angeles, 
Los Angeles, CA 90095, USA}}
\newcommand{\stonybrook}{\affiliation{C.N.Yang Institute for Theoretical Physics, State University of New York 
at Stony Brook, Stony Brook, NY 11794, USA}}
\newcommand{\beijing}{\affiliation{Institute of Physics, Chinese Academy of Sciences, Beijing 100080, China}}
\newcommand{\ucl}{\affiliation{Department of Physics and Astronomy, University College London, 
Gower Street, London WC1E 6BT, United Kingdom}}

\begin{document}
\title{Boundary effects to the entanglement entropy and two-site entanglement \\ of the spin-1 valence-bond solid}
\author{Heng Fan}\beijing
\author{Vladimir Korepin}\stonybrook
\author{Vwani Roychowdhury}\ucla
\author{Christopher Hadley}\ucl
\author{Sougato Bose}\ucl
\pacs{	75.10.Pq,	%
	03.67.Mn,	%
	03.65.Ud,	%
	05.70.Jk.	%
}
\date{\today}

\begin{abstract}
We investigate the von Neumann entropy of a block of subsystem for the valence-bond solid (VBS) state with general open boundary conditions.  
We show that the effect of the boundary on the von Neumann entropy decays exponentially fast in the distance between the subsystem 
considered and the boundary sites.  Further, we show that as the size of the subsystem increases, 
its von Neumann entropy exponentially  approaches the summation of the von Neumann entropies of the two ends, the exponent being related to the size. 
In contrast to critical systems, where boundary effects to the von Neumann entropy decay slowly, the boundary effects in a VBS, 
a non-critical system, decay very quickly. We also study the entanglement between two spins. 
Curiously, while the boundary operators decrease the von Neumann entropy of $L$ spins, they increase the entanglement between two spins. 
\end{abstract}
\maketitle


\section{Introduction}
Recently much research has been undertaken to understand the subtle interplay between {\it quantum entanglement} 
and {\it quantum criticality} for spin systems \cite{OAFF,ON,ABV,VLRK,K,VPC,FKR,VBR,JK}.  
Vidal \etal \cite{VLRK} showed that the entanglement between a block of contiguous spins and its complement 
in the ground state of the Ising model shows different behaviours for the gapped and gapless cases (critical and non-critical).  
The entanglement of the VBS ground state of the much-studied Affleck\ndash Kennedy\ndash Lieb\ndash Tasaki (AKLT) model 
\cite{AKLT,AKLT0,Affleck} 
is considered in Refs. \cite{VPC,FKR}, and very recently Campos Venuti \etal studied this state's long-distance 
entanglement property \cite{VBR}. The entanglement of the fermionic system was studied in Refs.\cite{WGK,GK}.

While much theoretical work in this area has focused on periodic boundary conditions, the open boundary 
condition has also attracted recent attention \cite{LSCA,ZBFS}.
Laflorencie \etal \cite{LSCA} numerically studied the boundary effects in the critical scaling of 
entanglement entropy (von Neumann entropy of a block of spins) for the (gapless) 1D XXZ model, 
and found the entanglement entropy {\it slowly} 
decays away from the boundary with a power-law. This result  
can be interpreted as stating that in critical systems, the boundary effects to the entanglement entropy 
is quasi-long-ranged; \ie there is a quasi-long-ranged entanglement between the boundaries and the subsystem 
in question.  This agrees with the fact that entanglement entropy increases logarithmically with the size of 
the subsystem in critical systems \cite{VLRK, CC}.  By contrast, the entanglement entropy for non-critical systems saturates 
to a constant bound when the subsystem size is increased, implying that the entanglement in the bulk is short-ranged. 
It is this localised nature of the entanglement entropy around the block edges which gives rise to an area law \cite{CC,PEDC} 
and makes ground states of gapped 1D systems particuarly amenable to simulation through matrix product states (MPS) \cite{MPSreview, Vidal03}.  

One might expect that boundary effects to the entanglement entropy also have different behaviours for 
critical and non-critical systems, and an interesting question then is whether the boundary effects to 
the entanglement entropy are short- or long-ranged, and what the exact behaviour of these are.  
This is one of the main motivations of this work: here we study the entanglement entropy of a VBS with general open boundary conditions. 
{\it We will show that, in contrast with the critical XXZ chain, the boundary effect to the entanglement entropy in the VBS state is short-ranged.} 
We will also show that the saturated bound for this state
is the sum of von Neumann entropy of the two boundaries.
Furthermore, it is not a constant (as is the case for a fixed boundary condition); 
it varies for different boundary conditions. This saturated bound in 1D state corresponds to the area 
law for higher-dimensional systems.

This model was originally studied by Affleck {\it et al.} in the context of the Haldane conjecture \cite{AKLT, AKLT0, Affleck}.  
It has also been the focus of much renewed interest since its generalisation to MPS\mdash which have been shown to efficiently 
simulate many 1D systems \cite{MPSreview, Vidal03} and may be used as a variational set in density matrix 
renormalisation group (DMRG) calculations\mdash \cite{S,Verstraete} and the discovery that its analogue in 2D is 
a resource for universal quantum computation \cite{VCb}.  
We hope that studying the boundary effect on the entanglement entropy will give some further insight into this model.
Since the boundary effects to the block entanglement entropy decays fast, we expect the
the DMRG method can be applied efficiently to this model.   

\section{Definition of the VBS state}
The spin-1 VBS state with general open boundary conditions (GOBC) \cite{TS} takes the form:
\begin{align}
|{\rm VBS}\rangle =Q_{l}^p
\left[\prod _{k=-N_l+1}^{L+N_r-1}(a^{\dag }_k b^{\dag}_{k+1}-b^{\dag }_k a^{\dag }_{k+1})\right]
Q_{r}^q|{\rm vac}\rangle\label{generalvbs}
\end{align}
where $a_k^{\dagger },b_k^{\dagger }$ are bosonic operators,
$Q_{l}^p$ and $Q_{r}^q$ are respectively the left and right boundary 
operators; $p,q=\pm $ with $Q_l^+=a_{-N_l+1}^{\dag }$, $Q_l^-=b_{-N_l+1}^{\dag }$, 
$Q_r^+=a_{L+N_r}^{\dag }$ and $Q_r^-=b_{L+N_r}^{\dag }$; $|{\rm vac}\rangle $ is 
the vacuum state; and $N_l,N_r$ are integer numbers. Since the left and right boundary operators are 
mutually independent, there are altogether four different VBS states with GOBC.  
Note that all sites in the spin chain including the left and right boundary sites $-N_l+1$, $L+N_r$ 
are spin-1's. Thus this VBS state (\ref{generalvbs}) is different from that studied in Ref. \cite{FKR}.  
We should also note one boundary operator, for example, $Q_l^+$ changes
the boundary state $a^{\dag }_{-N_l+1}|vac\rangle $ and 
$b^{\dag }_{-N_l+1}|vac\rangle $. 
In fact, the state (\ref{generalvbs}) is the ground state of the the Hamiltonian studied by Affleck \etal \cite{AKLT}, 
\begin{eqnarray}
{\cal {H}}=\sum_{j=-N_l+1}^{L+N_r-1}\left[ ({\bf S}_j\cdot {\bf S}_{j+1})+\frac {1}{3}
({\bf S}_j\cdot {\bf S}_{j+1})^2\right] . 
\end{eqnarray}

\section{Dividing the chain}
For convenience in later calculations, we divide this 1D state into three parts: 
the left-hand, central and right-hand parts. The left-hand part is defined as
$|{\rm left},p\rangle = Q_{l}^p 
\prod _{k=-N_l+1}^{0}(a^{\dag }_k b^{\dag}_{k+1}-b^{\dag }_k a^{\dag }_{k+1})
|{\rm vac}  \rangle $.
Similarly, the right-hand part is defined as
$|{\rm right},q\rangle =\prod _{k=L+1}^{L+N_r-1}(a^{\dag }_k b^{\dag
}_{k+1}-b^{\dag }_k a^{\dag }_{k+1})Q_{r}^q|{\rm vac}\rangle $.
Finally, the central part is written $|{\rm central}\rangle =\prod _{k=1}^{L}(
a^{\dag }_k b^{\dag}_{k+1}-b^{\dag }_k a^{\dag }_{k+1})
|{\rm vac}  \rangle $. Note that site 1 appears in both the left and central parts, 
and acts as two spin-1/2's; site $L$ similarly appears in both the central and right parts.  
Thus the whole VBS state with GOBC now takes the form 
$|{\rm VBS};p,q\rangle =|{\rm left},p\rangle |{\rm central}\rangle |{\rm right},q\rangle $.  
We should note that this is not strictly a product state, 
but that this decomposition is valid for our purposes.  
Double-counting is avoided since each bulk spin consists of two spin-1/2's and 
the two bosonic operators (spin-1/2's) in one site constitute a spin-1 state by Fock space
representation.  For example, terms $(a^{\dag }_0 b^{\dag}_{1}-b^{\dag }_0 a^{\dag }_{1})
|{\rm vac}  \rangle$ and  $(a^{\dag }_1 b^{\dag}_{2}-b^{\dag }_1 a^{\dag }_{2})
|{\rm vac}  \rangle$ belong to left and central parts, respectively, however,
by Fock space representation, the product state will create at site 1 the state
$(a^{\dag }_{1})^2|{\rm vac} \rangle , (b^{\dag }_{1})^2|{\rm vac} \rangle $ and
$a^{\dag }_{1}b^{\dag }_{1}|{\rm vac} \rangle $. Thus the three parts are connected
to constitute the original state (\ref{generalvbs}).
Our aim is to now study the von Neumann entropy of the reduced density 
operator of the contiguous spins from site 1 to $L$ of the state $|{\rm VBS};p,q\rangle $. 
For this aim, according to the theory of entanglement, 
the left- and right part states $|{\rm left},p\rangle $ and $|{\rm right},q\rangle $ 
can be replaced by two bipartite states, through the Schmidt decomposition.

Without loss of generality, we start from the left part and consider the entanglement of 
the quantum state $|{\rm left},p\rangle $ between site 1 and the rest; \ie we consider it a 
bipartite state with site $1$ as one particle and the rest as another particle.  
According to the Schmidt decomposition, we can first calculate eigenvalues of the reduced density 
operator of site 1 for state $|{\rm left},p\rangle $. 

\begin{figure}[ht]
\vskip 1.5truecm
\includegraphics[width=6cm]{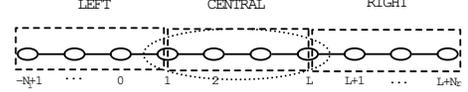}
\vskip 1truecm
\caption{The quantum spin chain with $L+N_l+N_r$ sites; each is spin-1. We calculate the 
von Neumann entropy of $L$ spins in the chain.
The whole spin chain is divided into three parts: the left, central and right parts. 
The spin-1 at site 1 is divided into two spin-1/2's; one in the left part, one in the central part.  
Similarly the spin-1 at site $L$ is split between the central- and right parts.}
\end{figure}

Denote $|\Psi ^{-}\rangle _{k,k+1}\equiv (a^{\dag}_k b^{\dag }_{k+1}-b^{\dag }_k a^{\dag }_{k+1})|{\rm vac}\rangle $, 
we know $|\Psi ^-\rangle _{01}|\Psi ^-\rangle _{12}\frac {1}{\sqrt{3}}\sum _{\alpha _k=1}^3|\alpha _1\rangle(I\otimes \sigma _{\alpha _1})
|\Psi ^-\rangle _{0,2}$,
where $\{\sigma_i\}_{i=0}^{3}$ is the Pauli group, and we have defined the 
states $\ket{\alpha_1}=I\otimes\sigma_{\alpha_1}\ket{\Psi^-}$.  
Here $\sigma _1=a_1^{\dag }b_1+a_1b_1^{\dag}$,
$\sigma _2=-\mi a_1^{\dag }b_1+\mi a_1b_1^{\dag }$, 
$\sigma _3=a_1^{\dag }a_1-b_1^{\dag }b_1$ and $\sigma_0 = a_1^{\dag}a_1 + b_1^\dag b_1$, the identity.
By this result, the state of the left part may be written
\begin{align}
|{\rm left},p\rangle 
&=\frac {1}{3^{(N_l-1)/2}}\sum _{\alpha _{0},\ldots,\alpha_{-N_l+2}=1}^3
|\alpha _{-N_l+2}\rangle \otimes\cdots\otimes|\alpha _0\rangle \nonumber \\
&\times (Q_l^p\otimes \sigma _{\alpha _0}\cdots\sigma _{\alpha_{-N_l+2}})|\Psi ^-\rangle _{-N_l+1,1}.
\end{align}
It is now possible to calculate the site 1 reduced density operator. 
Using the identity $\sum _{\alpha =1}^3(I\otimes \sigma _{\alpha })|\Psi ^-\rangle 
\langle \Psi ^-|(I\otimes \sigma _{\alpha })^{\dag }=I-|\Psi ^-\rangle 
\langle \Psi ^-|$ (where $I$ on the l.h.s. and r.h.s. is the identity in $\C^2$ and $\C^2\otimes\C^2$, respectively) 
we find
\begin{eqnarray}
\rho _1={\rm Tr_1}_{}(Q_l^p\otimes I)[\frac {1}{4}(1-f_l)I+f_l|\Psi ^- \rangle 
\langle \Psi ^-|](Q_l^p\otimes I)^{\dag },\label{site1}
\end{eqnarray}
where $f_l=(-\frac {1}{3})^{N_l-1}$, and the trace is over the first Hilbert space. 
Now we find that the matrix form of the reduced density operator of site 1 takes a diagonal form
$\rho _1={\rm diag}(\xi _l^+,\xi _l^-)$,
where we have defined $\xi^\pm_{l}=(3\pm f_{l})/3$ and $f_{l}=(-1/3)^{N_{l}-1}$.  
For different boundary operators $Q_l^{\pm }$, state $\rho _1$ is invariant under a basis transformation. 
By entanglement theory, we can replace the 
quantum state of left part by a real bipartite state
$|{\rm left},p\rangle \rightarrow
|\phi _l\rangle \equiv \left(\sqrt {\xi_l^+}a_0^{\dag }b_1^{\dag } - 
\sqrt {\xi_l^-}b_0^{\dag }a_1^{\dag }\right)|{\rm vac} \rangle $.
We find that the reduced density operator $\rho _1$ converges to the identity exponentially fast with respect to $N_l$, 
and thus we can simply consider $|\phi _l\rangle $ as a singlet 
state $|\Psi^-_{0,1}\rangle $ when $N_l\rightarrow \infty $. 
Similarly for right part of the state, we have
$|\phi _r\rangle \equiv \left(\sqrt{\xi_r^+}a_L^{\dag }b_{L+1}^{\dag } - 
\sqrt{\xi_r^-}b_L^{\dag }a_{L+1}^{\dag }\right)|\rm{vac} \rangle$,
where the $\xi_r^{\pm}$ have a similar definition to the $\xi_l^\pm$.  Thus the VBS state in Eq. (\ref{generalvbs}) 
may be rewritten
\begin{align}
|{\rm VBS}\rangle = |\phi _l\rangle \prod _{k=1}^{L-1}(a^{\dag }_k b^{\dag
}_{k+1}-b^{\dag }_k a^{\dag }_{k+1})|\phi _r\rangle 
 |{\rm vac}\rangle ,
\label{state2}
\end{align} 
where indices $p,q$ are suppressed since they do not change the result.
The validity of the transformation from  
(\ref{generalvbs}) to (\ref{state2}) in studying the von Neumann 
entropy of contiguous $L$ spins can also be checked by a 
method with matrix product state representation introduced 
in, for example, Ref.\cite{Verstraete}. A numerical calculation 
for small $L$ confirms our result.

\begin{figure}[ht]
\vskip 1truecm
\includegraphics[width=4cm]{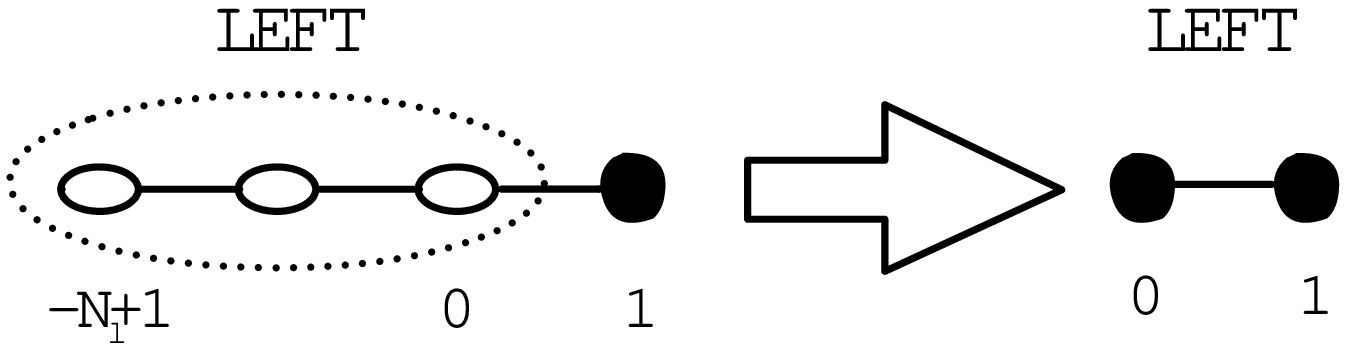}
\hskip 0.6truecm
\includegraphics[width=3cm]{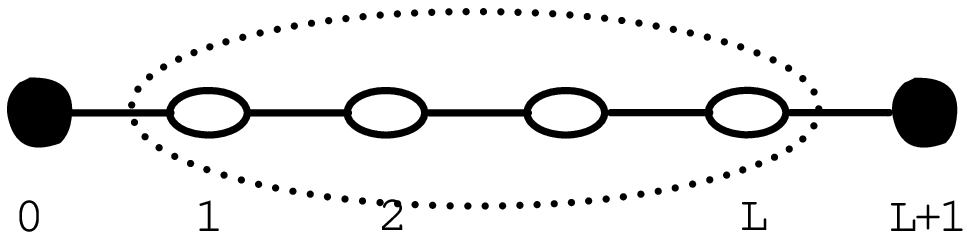}
\vskip 1truecm
\caption{The quantum state of left part is considered as a bipartite state with one particle 
in site 1 and the rest as another particle. According to the Schmidt decomposition this state is 
equal to a bipartite state $|\phi _l\rangle $, 
and each spin is spin-$1/2$ at site 0 and 1;
Finally, the VBS state with GOBC is mapped to a 
state with $L+2$ sites with two spin-1/2 boundaries at ends $0$ and $L+1$}
\end{figure}

In the case that $N_l\rightarrow\infty$, $N_r\rightarrow\infty$ the state takes 
the form $|{\rm VBS}\rangle = \prod _{k=0}^L(a^{\dag }_k b^{\dag}_{k+1}-b^{\dag }_k a^{\dag }_{k+1})|{\rm vac}\rangle $.  
The entanglement entropy of this state has been studied previously \cite{FKR}; it was found that there are no boundary effects.   
In this paper, one of our main concerns is to show that the VBS state (\ref{generalvbs}) {\it does} 
have a boundary effect. Already we know that if the block of $L$ contiguous spins in Eq. (\ref{generalvbs}) 
is far from the two boundary sites the boundary effect will decay very rapidly (exponentially).  
We now present explicitly the entanglement entropy of these $L$ spins.  
Let us first rewrite the left part state in the form
$|\phi _l\rangle =(V_l\otimes I)|\Psi ^-\rangle _{01}$,
where the matrix form of $V_l$ takes the form $V_l= {\rm diag}(\xi ^+_l,\xi _l^-)$. 
Note that $V_l$ is not necessarily unitary. 
In a similar manner for right part state ,  $V_r$ can also be defined $V_r={\rm diag}(\xi _r^-,\xi _r^+)$, 
giving $|\phi _r\rangle  =(I\otimes V_r)|\Psi ^-\rangle _{L,L+1}$.  
We can then write the VBS with GOBC in the form:
\begin{eqnarray}
|{\rm VBS}\rangle 
&=& (V_l\otimes V_r)_{0,L+1}\sum_{\{\alpha_i\}} 
|\alpha _1\rangle\otimes\cdots\otimes|\alpha _L\rangle \nonumber \\
&&\times (\sigma _{\alpha _1}\cdots\sigma _{\alpha _L}\otimes I)|\Psi ^-\rangle _{0,L+1},
\end{eqnarray}
where the summation is from 1 to 3 for indices  $\alpha _1,\cdots,\alpha _L$. According to entanglement theory, 
the von Neumann entropy of the reduced density operator
of $L$ spins is the same as the von Neumann entropy of the two ends (sites $0$ and $L+1$). 
The reduced density operator of these sites takes the form
\begin{align}
\tilde{\rho }_L 
= (V_l\otimes V_r) \left\{\frac {1-p}{4}I+p|\Psi ^-\rangle \langle \Psi ^-|\right\}(V_l\otimes V_r)^{\dag}
\end{align}
where $p=(-1/3)^{L}$. 
\begin{widetext}
Expanding this, we find the following form for the density matrix:
\begin{align}
&\tilde {\rho }_L = \nonumber\\
&\left(
\begin{array}{cccc}
\frac {1-p}{4}(\xi _l^+\xi _r^-)^2 & 0 & 0 & 0 \\
0& \frac {1+p}{4}(\xi _l^+\xi _r^+)^2 &  -\frac {p}{2}\xi _l^+\xi _l^-\xi _r^+\xi _r^- & 0\\
0 & -\frac {p}{2}\xi _l^+\xi _l^-\xi _r^+\xi _r^- &  \frac {1+p}{4}(\xi _l^-\xi _r^-)^2 &0\\
0 & 0 & 0 & \frac{1-p}{4} (\xi^-_l\xi^+_r)^2
\end{array}\right)\nonumber 
\end{align}
When $p$ is small the eigenvalues of the matrix $\tilde{\rho} _L$ can be found by a Taylor expansion to be
$\lambda _1=\xi_l^+\xi_r^-(1-p)$; $\lambda _2=\xi_l^-\xi_r^+(1-p)$;
$\lambda _3=\xi_l^+\xi_r^+(1+p)+O(p^2)$; $\lambda _4=\xi_l^-\xi_r^-(1+p)+O(p^2)$.
Recall that $|\phi _l\rangle $ is a pure state, so the reduced density 
operators at sites 0 and 1 are the same under unitary transformation (\ie $\rho _0 = \rho _1)$; 
similarly for state $|\phi _r\rangle $. By checking the eigenvalues of $\tilde {\rho }_L$, 
we find that $\tilde {\rho }_L=\rho _0\otimes \rho _{L+1}+O(p)$, where the equation is true under a unitary transformation.  
This transformation has no effect on the von Neumann entropy, and thus we suppress it.  The density operator of $\tilde {\rho }_L$ 
converges exponentially fast to the 
tensor product of two ends; the speed is $p=(-1/3)^L$. 
\end{widetext}

\section{Block entropy}
Finally we calculate explicitly the von Neumann 
entropy of a block of $L$ spins,
\begin{eqnarray}
S(\tilde {\rho }_L)=S(\rho _0)+S(\rho _{L+1})+O(p).\label{ourresult}
\end{eqnarray}
The von Neumann entropy of $L$ spins converges to the 
von Neumann entropy of two ends exponentially fast with 
convergence $p=(-1/3)^L$. The exact form of $O(p)$ is written below 
\footnote{$O(p)=p [ S(\rho _0)+S(\rho _{L+1})-4f_lf_r
+ (\xi_l^+ \xi_r^-/2)\log_2 (\xi_l^+\xi_r^-/4) + (\xi_l^- \xi_r^+/2)\log_2 (\xi_l^-\xi_r^+/4)]$}. 
For a pure bipartite state $|\phi _l\rangle $,
we know $S(\rho _0)=S(\rho _1)=-\frac {3-f_l}{6}\log \frac {3-f_l}{6} -\frac {3+f_l}{6}\log \frac {3+f_l}{6}$. 
When $f_l$ is small, we find $S(\rho _0)\sim 1-\frac {f_l^2}{18}$
and similarly for $S(\rho _{L+1})$.
So when the distances between the block of $L$ contiguous spins and the two ends are large, we know that
\begin{eqnarray}
S(\tilde {\rho _L})\sim 2-\frac {f_l^2+f_r^2}{18}+O(p)+O(f_l^4)+O(f_r^4).
\label{entropy}
\end{eqnarray}
Note that when $p$ is comparable with $f_l$ and $f_r$, then $O(p)$ may 
have contributions for both $f_l$ and $f_r$.  Considering all of these points, we conclude that boundary 
effects decay exponentially when the distances between the subsystem and the boundaries increase. 
In the case that there are no boundary operators, the trace in Eq. (\ref{site1}) is over the identity, 
and $\rho _1$ reduces to $I$.  Hence the boundary effects to the entanglement entropy never arise; 
this is the case previously studied \cite{FKR}.  
We remark for 1D VBS with GOBC,  
the terms corresponding to the topological entropy in Ref.\cite{KPLW,LW} are the boundary 
terms appeared in Eq.(\ref{entropy}), and they remain unchanged for $L\rightarrow \infty $. 

\section{Comparison with a critical system}
It was found numerically \cite{LSCA} that for spin-1/2 XXZ chains the entropy takes the form
$S(L,N)=S_U(L,N)+(-1)^LS_A(L,N)$, where the second term arises from the boundary conditions,
and can be written $S_A(L,N)=1/(\sin (2\pi N_{n,r}/N)N/\pi )^K$. 
Note that the notation of Ref. \cite{LSCA} has been changed slightly here and $N=N_r+N_l+L$, $N_l=N_r=N_{n,r}$; 
$K$ depends on
the anisotropy parameter in the XXZ chains and $K=1$ for an XX chain. We find that the
boundary term decays slowly and is quasi-long-ranged, while for the VBS studied
in this paper, the boundary terms in Eq. (\ref{entropy}) decay exponentially fast.

\section{Two-site entanglement by negativity and realignment calculations}
We next study the boundary effects to entanglement
between only {\it two spins} in the bulk. 
Consider spins $1$, $L$. The previous method still works; \ie we can reduce the length of the chain
from $N_l+L+N_r$ sites in Eq. (\ref{generalvbs}) to a chain with only $L+2$ spins in Eq. (\ref{state2}). 
This transformation does not change the entanglement between two spins at sites 1 and $L$.  
We find that the density operator of the bipartite state is written
\begin{align}
&\rho _{1,L}=\sum |\alpha _1\alpha _L\rangle \langle 
\alpha _1'\alpha _L'|\,{\rm Tr}\,(V_l\otimes V_r)(\sigma _{\alpha _1}\otimes \sigma _{\alpha _L}^t)\nonumber \\
&\times\left[ \frac {1-p}{4}I+p|\Psi ^-\rangle \langle \Psi ^-|\right](\sigma _{\alpha '_1}\otimes 
\sigma _{\alpha '_L}^t) ^{\dagger }(V_l\otimes V_r)^\dag
\end{align}
The explicit form of this density matrix is complicated. However since
the boundary matrices $V_l$ and $V_r$ become identities exponentially with
$N_l$ and $N_r$, we know that the boundary effects
to this density operator decay exponentially if the separation of these two spins with the boundaries
increase. Thus the case that $N_r$ or $N_l$ is small already can provide enough information about the
entanglement between the considered two spins. The cases $N_r\to\infty $ and
$N_l\to\infty $ are special since they correspond to the cases that there are no 
boundary operators $Q_r^{\pm }$ and $Q_l^{\pm }$, respectively.

In order to quantify the two-spin entanglement in this case, we shall use {\it negativity}, $\mathcal{N}$ \cite{VW}.  
We find that there is no entanglement for non-nearest neighbouring spins (regardless of boundary).  
The nearest-neighbour negativity for a range of $N_l$, $N_r$ is presented in Table \ref{negativitytable}, 
where a factor $1/9$ is omitted.  

\begin{table}
\begin{tabular}{|c|c|c|c|c|c|}
\hline
&$N_r=1$  &$N_r=2$  & $N_r=3$ & $N_r=4$  & $N_r=\infty $ \\
\hline
$N_l=1$ &1.45919  &1.50111  &1.43456  &1.45142 & 1.44170\\
\hline
$N_l=2$ &1.50111  &1.05433  &1.15504  &1.11552 & 1.12486\\
\hline
$N_l=3$ &1.43456  &1.15504  &1.00609  &1.05018 & 1.03861\\
\hline
$N_l=4$ &1.45142  &1.11552  &1.05018  &1.00068 & 1.01252\\
\hline
$N_l=\infty$ &1.44670  &1.12486  &1.03861  &1.01252 & 1 (exact)\\
\hline
\end{tabular}
\caption{Nearest-neighbour negativity}
\label{negativitytable}
\end{table}

\begin{table}
\begin{tabular}{|c|c|c|c|c|c|}
\hline
&$N_r=1$  &$N_r=2$  & $N_r=3$ & $N_r=4$  & $N_r=\infty $ \\
\hline
$N_l=1$ &0.37393  &0.23445  &0.19692  &0.20032 & 0.19861\\
\hline
$N_l=2$ &0.23445  &0.03974  &0.02631  &0.02194 & 0.02221\\
\hline
$N_l=3$ &0.19692  &0.02631  &0.00439  &0.00293 & 0.00247\\
\hline
$N_l=4$ &0.20032  &0.02194  &0.00293  &0.00049 & 0.00027\\
\hline
$N_l=\infty$ &0.19861  &0.02221  &0.00247  &0.00027 & 0 (exact)\\
\hline
\end{tabular}
\caption{Nearest-neighbour entanglement by realigment}
\label{realignmenttable}
\end{table}

Curiously, we see that while boundary operators {\it decrease} the block 
von Neumann entropy{\mdash}see Eq.(\ref{ourresult}){\mdash}they {\it increase} the nearest-neighbour negativity.  
Roughly, this can be understood that the entanglement is monogamous \cite{CKW,OV0}, 
i.e., it can not be shared freely by many parties. 
A simple example about the monogamy of entanglement is that, suppose A, B and C are three parties,
if A and B are maximally entangled, A and C will be separable. 
The result in this paper shows that the boundary operators have effect on the
entanglement sharing in this many-body system. 
 
In the table above, we see that for finite $N_{l,r}$, it is always that case that $\mathcal{N}>1/9$.  
Asymmetric boundary operators may also increase the entanglement.  
For example, $\mathcal{N}=1.45919/9$ for $N_l=N_l=1$, but $\mathcal{N}=1.50111/9$ for $N_l=1,N_r=2$.  
Now $\mathcal{N}$ does not vary monotonically with $N_{l,r}$; \eg when $N_r$ varies from 
1 to $\infty$ (for $N_l=1$), $\mathcal{N}$ oscillates.  This seemingly new result 
may potentially be useful: nearest-neighbour entanglement may be controlled by tweaking boundaries.  

The drawback of the {\it negativity} is that the
bound entanglement cannot be detected and quantified. A complementary quantity derived from
the realignment separability criterion can partially solve this problem \cite{CAF}. 
For a two-site density operator $\rho $,
this quantity is defined as $\mathcal{R}=(||R(\rho )||-1)/2$, where the matrix $R(\rho )$ is obtained from
the density operator $\rho $ by the realignment method, and $||\cdot||$ is the trace norm. 
The larger of the quantities $\mathcal{N}$ and $\mathcal{R}$ gives a lower bound of the {\it concurrence} $\mathcal {C}$ 
for a mixed state in arbitrary dimensional systems; \ie $\mathcal {C}\ge {\rm max}\{ \mathcal{N}, \mathcal{R}\} $ \cite{CAF}.

Using this measure, we still find zero entanglement between detected for non-nearest neighbouring spins regardless of boundary.
The entanglement of nearest neighbouring spins by realignment is presented in 
Table \ref{realignmenttable} and
a factor $1/9$ is also omitted. We also still find that the boundary operators increase entanglement
$\mathcal{R}$, and indeed no entanglement is found for the case without boundary operators: this is due to the 
limitation of the realignment method. Finally let us remark that for the model studied in this paper, 
the {\it negativity} provides a stronger lower bound for the {\it concurrence}. Since the concurrence
for a general mixed state in ${\cal {C}}^3\otimes {\cal {C}}^3$ is difficult to find, the entanglement measures
by {\it negativity} and {\it realignment} are widely accepted.

\section{Conclusion}
In summary, we studied the boundary effects to the entanglement
entropy and the two-site entanglement for the spin-1 VBS state which is the
ground state for a gapped model. We showed that the
boundary effects are short-ranged, \ie they decays exponentially in the distance
between the subsystem considered and the boundary sites. This is different
from the case of XXZ chain which is a gapless model.
The two-site entanglement was studied by two entanglement measures,
the negativity and the realignment method. For the VBS state, 
we find the boundary operators decrease
the block von Neumman entropy but increase the  the nearest-neighbour 
entanglement measured by negativity and realignment.


\section{Acknowledgments}
H.F. was supported by `Bairen' program, NFSC and `973' program (2006CB921107), 
and V.K. was supported in part by NSF Grant DMS 0503712. C.H. and S.B. acknowledge UK EPSRC grants EP/P500559/1 and GR/S62796/01.  
We thank F. Verstraete for useful discussions.

\end{document}